# Democratising Knowledge Representation with BioCypher

Sebastian Lobentanzer[1], Patrick Aloy[2,3], Jan Baumbach[4], Balazs Bohar[5,6], Pornpimol Charoentong[7,8], Katharina Danhauser[9], Tunca Doğan[10,11], Johann Dreo[12,13], Ian Dunham[14,15], Adrià Fernandez-Torras[2], Benjamin M. Gyori[16], Michael Hartung[4], Charles Tapley Hoyt[16], Christoph Klein[9], Tamas Korcsmaros[5,17,18], Andreas Maier[4], Matthias Mann[19,20], David Ochoa[14,15], Elena Pareja-Lorente[2], Ferdinand Popp[21], Martin Preusse[22], Niklas Probul[4], Benno Schwikowski[12], Bünyamin Sen[10,11], Maximilian T. Strauss[19], Denes Turei[1], Erva Ulusoy[10,11], Judith Andrea Heidrun Wodke[23], Julio Saez-Rodriguez[1,*]

[1] Heidelberg University, Faculty of Medicine, and Heidelberg University Hospital, Institute for Computational Biomedicine, Bioquant, Heidelberg, Germany

[2] Institute for Research in Biomedicine (IRB Barcelona), the Barcelona Institute of Science and Technology, Barcelona, Catalonia, Spain

[3] Institució Catalana de Recerca i Estudis Avançats (ICREA), Barcelona, Catalonia, Spain

[4] Institute for Computational Systems Biology, University of Hamburg, Germany

[5] Earlham Institute, Norwich, UK

[6] Biological Research Centre, Szeged, Hungary

[7] Centre for Quantitative Analysis of Molecular and Cellular Biosystems (Bioquant), Heidelberg University, Im Neuenheimer Feld 267, 69120, Heidelberg, Germany

[8] Department of Medical Oncology, National Centre for Tumour Diseases (NCT), Heidelberg University Hospital (UKHD), Im Neuenheimer Feld 460, 69120, Heidelberg, Germany

[9] Department of Pediatrics, Dr. von Hauner Children's Hospital, University Hospital, LMU Munich, Germany

[10] Biological Data Science Lab, Department of Computer Engineering, Hacettepe University, Ankara, Turkey

[11] Department of Bioinformatics, Graduate School of Health Sciences, Hacettepe University, Ankara, Turkey

[12] Computational Systems Biomedicine Lab, Department of Computational Biology, Institut Pasteur, Université Paris Cité, Paris, France

[13] Bioinformatics and Biostatistics Hub, Institut Pasteur, Université Paris Cité, Paris, France

[14] European Molecular Biology Laboratory, European Bioinformatics Institute (EMBL-EBI), Wellcome Genome Campus, Hinxton, Cambridgeshire CB10 1SD, UK

[15] Open Targets, Wellcome Genome Campus, Hinxton, Cambridgeshire CB10 1SD, UK

[16] Laboratory of Systems Pharmacology, Harvard Medical School, Boston, USA

[17] Imperial College London, London, UK

[18] Quadram Institute Bioscience, Norwich, UK

[19] Proteomics Program, Novo Nordisk Foundation Centre for Protein Research, University of Copenhagen, Copenhagen, Denmark

[20] Department of Proteomics and Signal Transduction, Max Planck Institute of Biochemistry, Martinsried, Germany

[21] Applied Tumour Immunity Clinical Cooperation Unit, National Centre for Tumour Diseases (NCT), German Cancer Research Centre (DKFZ), Im Neuenheimer Feld 460, 69120, Heidelberg, Germany

[22] German Centre for Diabetes Research (DZD), Neuherberg, Germany

[23] Medical Informatics Laboratory, University Medicine Greifswald, Germany

* Corresponding author: pub.saez@uni-heidelberg.de

All authors except first and last are listed alphabetically

# Abstract

Standardising the representation of biomedical knowledge among all researchers is an insurmountable task, hindering the effectiveness of many computational methods. To facilitate harmonisation and interoperability despite this fundamental challenge, we propose to standardise the framework of knowledge graph creation instead. We implement this standardisation in BioCypher, a FAIR (findable, accessible, interoperable, reusable) framework to transparently build biomedical knowledge graphs while preserving provenances of the source data. Mapping the knowledge onto biomedical ontologies helps to balance the needs for harmonisation, human and machine readability, and ease of use and accessibility to non-specialist researchers. We demonstrate the usefulness of this framework on a variety of use cases, from maintenance of task-specific knowledge stores, to interoperability between biomedical domains, to on-demand building of task-specific knowledge graphs for federated learning. BioCypher (https://biocypher.org) frees up valuable developer time; we encourage further development and usage by the community.

# Graphical Abstract

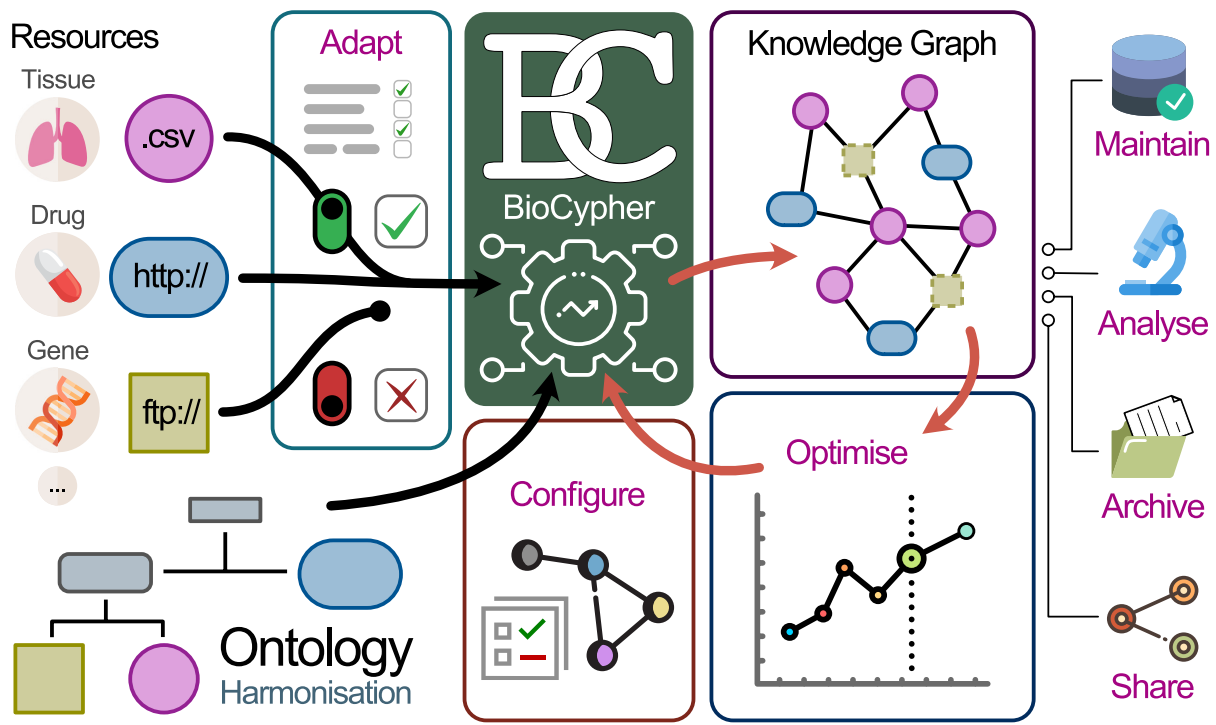

**Main Text**

**Introduction**

Biomedical knowledge, although increasingly abundant, is fragmented across hundreds of resources. For instance, a clinical researcher may use protein information from UniProtKB [1], genetic variants from COSMIC [2], protein interactions from IntAct [3], and information on clinical trials from ClinicalTrials.gov [4]. Combining these complementary datasets is a fundamental requirement for exhaustive biomedical research and thus has motivated a number of integration efforts to form harmonised knowledge graphs (KGs; i.e., knowledge representations based on a machine-readable graph structure). Decisions made on how to represent the knowledge at each primary source pose many real-world problems in their recombination, for instance via the use of different identifier namespaces, levels of granularity, or licences [5,6]. However, directly standardising the representation of biomedical knowledge is not appropriate for the diverse research tasks in the community; there is no one-size-fits-all [5–8].

This heterogeneity directly affects the FAIRness of knowledge representation:

**Findability**: Since many KGs have been created, finding the one most suitable for a specific task is challenging and time-consuming [5,6].

**Accessibility**: Few available KG solutions perfectly fit the task the individual researcher wants to perform. Creating custom KGs is only possible for those that can afford years of development time by an individual [7,9] or even entire teams [10]. Smaller or non-bioinformatics labs need to choose from publicly available KGs, limiting customisation and the use of non-public data. There exist frameworks to build certain kinds of KG from scratch [8,11], but these are difficult to use for researchers outside of the ontology subfield and often have a rigid underlying data model [6,12]. Even task-specific knowledge graphs sometimes need to be built locally by the user due to licensing or maintenance reasons, which requires significant technical expertise [13].

**Interoperability**: For the above reasons, many KGs (Supplementary Table 1) are built manually for specific applications, which is very laborious and often redundant, since the primary data sources overlap substantially [5]. For downstream users, the resulting KGs are too distinct to easily compare or combine [6].

**Reusability**: Maintaining KGs for the community is additional work; once maintenance stops, they quickly deteriorate, leading to reusability and reproducibility issues [5]. Modifying an existing, comprehensive KG for a specific purpose is a non-trivial and often manual process prone to lack of reproducibility[14].

## Approach

To address these problems, we present BioCypher, a software that improves biomedical knowledge representation by means of:

1) **Modularity**: To facilitate the maintenance of multiple task-specific KGs from overlapping primary resources, we propose a modular approach that allows recombining individual data "adapters" for primary resources in a reusable manner. This allows delegating the maintenance work to one central place for each adapter instead of having to maintain the primary resource inside each individual KG (see **case study "Modularity"**).

2) **Harmonisation**: To facilitate harmonisation of datasets from a biological perspective, we propose to use ontology mapping (referring to a hierarchical organisation of biological concepts). Primary data sources may represent similar data in different ways. BioCypher harmonises biomedical data by mapping divergent representations onto the same biological concept, flexibly dealing with multiple technical implementations of ontology (see **case study "Tumour board"**). In addition, the ontological information projected onto each KG entity allows for more flexible and informative queries in downstream analyses (see **case study "Network expansion"**).

3) **Reproducibility**: By sharing the ontology mapping from (2) in a project-specific manner, a database used for a specific task can be reproduced more effectively. Since sharing the databases themselves is often prohibited by their large size, BioCypher facilitates the creation of task-specific subsets of databases to be shared alongside analyses. Extensive automation reduces development time and file sizes, while additionally making the shared dataset independent of database software versions (see **case studies "Network expansion", "Subgraph extraction", and "Embedding"**).

4) **Reusability and accessibility**: Finally, the sustainability of research software is strongly related to adoption in - and contributions from - the community. BioCypher is developed as open source software applying modern methods of continuous integration and deployment, including a diverse community of researchers and developers from the beginning (see **case study "Data integration"**). This facilitates robust workflows that are tested end-to-end, including the integrity of the scientific data. We operate under the permissive MIT licence and provide contributors with guidelines for their contributions and a code of conduct. To increase accessibility to the community, we create user-friendly interfaces using open standards (**Figure S5**). These interfaces, together with the biological perspective introduced by ontology mapping, improve usability by non-bioinformaticians.

Taken together, these features will make the process of gathering and harmonising biomedical knowledge simpler, more democratic, and FAIR [15].

## Implementation

We build on recent technological and conceptual developments in biomedical ontologies that greatly facilitate the harmonisation of biomedical knowledge. We integrate a consistent and comprehensive biomedical ontology, the Biolink model [16], and an extensive catalogue and resolver for biomedical identifier resources, the Bioregistry [17]. Both projects, like BioCypher, are open-source and community-driven. Biolink serves as a basis for the representation of biomedical concepts, and Bioregistry provides consistent vocabularies for these concepts as well as validation of identifiers. We also facilitate exchange, extension, and modification of the ontological scaffold to accommodate database-specific needs. BioCypher is implemented as a Python library that provides a low-code access point to data processing and ontology manipulation (for examples, see **case studies "Tumour board" and "Network expansion"**). BioCypher facilitates the decision on how to represent knowledge and simplifies the creation of the corresponding KG, bridging the gap between the field of biomedical ontology and the broad application of databases to biological research questions.

BioCypher's translation framework simplifies the creation of custom KGs through a combination of adapters (data ingestion) and a schema configuration (graph structure and ontology mappings). Building a task-specific KG, given existing configuration, takes only minutes, and creating a KG from scratch can be achieved in a few days of work. This allows for rapid prototyping and automated machine learning (ML) pipelines that iterate the KG structure to optimise predictive performance; for instance, building custom task-specific KGs for graph embeddings and ML (see **case study "Embeddings"**). Despite its speed, automated end-to-end testing of millions of entities and relationships per KG increases trust in the consistency of the data (see Methods for details and the **case study "Network expansion"** for an example).

By abstracting the KG build process as a combination of modular *input* adapters, BioCypher saves developer time in maintaining integrative resources made up of overlapping primary sources (**Figure 1A**, see **case study "Modularity"**). Several of these integrative resources are migrated into the BioCypher framework, for instance OmniPath [18,19], the Clinical Knowledge Graph (CKG [20]), CROssBAR v2 [21], the Bioteque [7], and a Dependency Map KG [22]. By mapping each of these knowledge collections onto the same ontological framework, we also gain automatic interoperability between the different biomedical domains (**Figure 1B**).

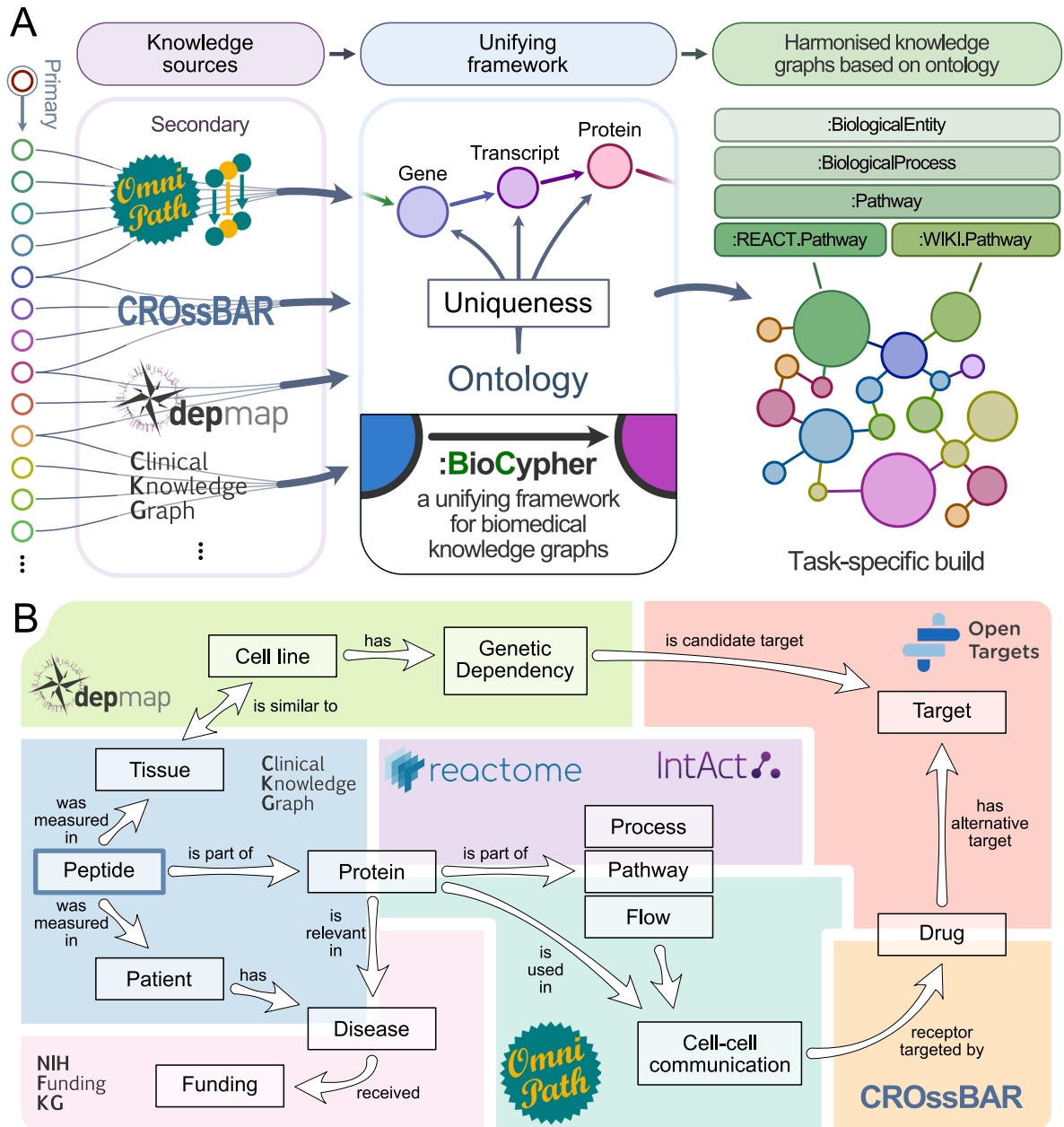

**Figure 1: The BioCypher framework.** A) We transform commonly used, curated resources into configurable, task-specific knowledge graphs, using ontology to inform biological "objectness", facilitating integration, reasoning, and interpretation. Later, the framework will support all primary knowledge sources in a modular fashion (see Figure S1). B) Agreeing on a common representational framework allows recombination of task-specific data sources to answer complex queries across biomedical domains. For instance, starting at mass spectrometry measurements of a patient's tumour (left), one could go through clinical annotations to genetic dependencies from the Dependency Map project to identify potential drug targets, or through pathway / process annotations in Reactome and IntAct, identify relevant ligand-receptor pairs using OmniPath, and use CROssBAR to perform drug discovery or repurposing for these receptors.

By providing modular *output* adapters, we can adjust to the various needs of KG users. A Neo4j adapter provides rapid access to extensive databases for querying from analysis (Jupyter) notebooks and facilitates maintenance of large knowledge collections for storage. A CSV writer allows exchange with other knowledge curation services, for instance in KGX format [12]. Python-native adapters (for instance to sparse matrix or NetworkX format) yield knowledge representations that can immediately be used programmatically in machine learning frameworks such as PyTorch Geometric for deep learning [23].

For high performance, we implement property graph database technologies that provide intuitive query interfaces, such as the Cypher graph query language developed by Neo4j [24]. This enables complex and versatile queries that pave the way towards rich and highly interactive interfaces. For example, web widgets and apps (such as drug discovery and repositioning with https://crossbar.kansil.org and analysis workflows with https://drugst.one) allow non-computational researchers to browse and customise the database, and to plug it into standard pipelines [25]. Additionally, a structured knowledge representation facilitates connection to modern natural language processing applications such as GPT [26]. Neo4j is highly scalable and interacts well with other components of large-scale, distributed, high performance computing infrastructure (see **case studies "Open Targets" and "Upscaling"**). Thanks to common standards, tools can be shared across projects and used community-wide or in cloud-based services that preserve sensitive patient data (see **case study "Federated learning"**).

## Discussion

There have been numerous attempts at standardising knowledge graphs and making biomedical data stores more interoperable [5,6]. They can be divided into three broad classes representing increasing levels of abstraction of the KG build process:

**i) Centrally maintained databases** include task-oriented data collections such as OmniPath [18] or the CKG [20]. They are the least flexible form of knowledge representation, usually bound to a specific research purpose, and are highly dependent on their primary maintainers for continuous functioning. BioCypher reduces the development and maintenance overhead that usually goes along with such a resource, making a task-specific KG feasible for smaller and less bioinformatics-focused groups. These databases usually do not conform to any standard in their knowledge representation, hindering their integration. In contrast, with BioCypher, we migrated OmniPath, CKG, and other popular databases onto an interoperable KG framework.

**ii) Explicit standard formats or modelling languages** include the Biolink model [16], BEL [27], GO-CAM [28], SBML [29], BioPAX [30], and PSI-MI [31]. There are many more, each a solution to a very specific problem, as reviewed elsewhere [27,32]; some are part of the COMBINE standard ecosystem [33]. Their main shortcoming is the rigidity that follows from their data model definitions: to represent data in one of these languages, the user needs to fully adopt it. If the task exceeds the scope of the language, the user needs to either look for alternatives, or introduce new features into the language, which can be a lengthy process. In addition, some features may be incompatible, and thus, one centrally maintained language definition is fundamentally limited. With BioCypher, each of the above languages can be adopted as the basis for a particular knowledge graph; in fact, we use the Biolink model as a basic ontology. Inside our framework, these languages can be freely and transparently exchanged, modified, extended, and hybridised, as we show in several of our case studies (e.g., **"Tumour board"** extends Biolink with Sequence Ontology).

**iii) KG frameworks** provide a means to build KGs, similar to the idea of BioCypher [8,11,12,34]. However, most tie themselves tightly to a particular standard format or modelling language ecosystem, thereby inheriting many of the limitations described above. The Knowledge Graph Hub provides a data loader pipeline, KGX allows conversion of KGs between different technical formats, and RTX-KG2 builds a fixed semantically standardised KG; all three adhere to the Biolink model [12,34]. Bio2BEL is an extensive framework to transform primary databases into BEL [11]. Finally, PheKnowLator is a framework that can use different languages, similar to BioCypher in that regard [8]. However, being rooted in Semantic Web technology, it is directed at knowledge representation experts, requiring considerable bioinformatics and ontology expertise.

## Conclusion

Biomedical knowledge is amassed at an ever-increasing rate, and machine learning tools that leverage prior knowledge in combination with biomedical big data are gaining much traction, yielding, for instance, sophisticated deep neural architectures that perform prediction of combinatorial perturbations or attempt to diagnose rare diseases [7,35–40]. However, the knowledge representations used in these frameworks result from arbitrary decisions about inclusion and structure, followed by manual implementation, and thus are neither optimised for the task at hand nor tested for alternatives or robustness with regard to the representation. BioCypher provides a timely framework for KG standardisation to improve the interoperability of prior knowledge sources and downstream computational analysis methods. We facilitate FAIRness in knowledge representation by increasing accessibility for non-bioinformatics groups and smaller labs and demonstrate the key advantages of BioCypher by examples in the Supplementary Materials.

Despite many efforts towards the standardisation of knowledge representation in biomedicine, there is no widely accepted solution. Very often, resources take the "path of least resistance" in adopting their own, arbitrary formats of representation. To our knowledge, there is no framework that provides easy access to state-of-the-art KGs to the "average" biomedical researcher, a gap that Biocypher aims to fill. We believe that creating a more interoperable biomedical research community is as much a social effort as it is a scientific software problem. To allow adoption of any paradigm, the process of adoption must be made as simple as possible, and adoption must yield tangible rewards, such as significant savings in developer time. We invite all database and tool developers to join our collective effort.

## Methods

BioCypher is implemented as a Python package. Its main purpose is to receive arbitrarily structured biomedical information and create or update a knowledge graph (KG) that unambiguously maps each KG entity to its corresponding biological class. It uses ontology (the curation of biomedical concepts into a hierarchy of classes) to encode semantic information about KG entities ("objectness") in the biomedical space, which is required to enable machine readability and automation. BioCypher can be described as an extract-transform-load pipeline with a focus on interoperability in biomedicine.

As a basis for representation of biomedical concepts, we use the Biolink model [16], a comprehensive and generic biomedical ontology; where needed, this ontology can be exchanged with or extended by more specific and task-directed ontologies, for instance from the OBO Foundry [41]. Identifier namespaces are collected from the community-curated and frequently updated Bioregistry service [17], which is important for ensuring continued compatibility among the created KGs. Bioregistry also supplies convenient methods for parsing identifier Compact URIs (CURIEs), which are the preferred method of unambiguously specifying identities of KG entities. For identifier mapping, where it is needed, the corresponding facilities of PyPath [18] are used and extended.

The preferred way of entering data into a BioCypher graph attaches scientific provenance to each entry, allowing the aggregation of data with respect to their sources (for instance, the publication an interaction was derived from) and thus avoiding problems such as duplicate counting of the same primary data from different secondary curations. In this way, confidence about knowledge contents of each graph can be assessed more easily, for instance in the order of "multiple curated sources" > "single curated source" > "multiple experimental sources" > "single experimental source" > "predicted interaction". For author attribution, the preferred way of entering data into BioCypher also includes the exact provenance of each entry, for instance, the publication it was derived from or the consortium responsible for the curation of said content. In the same way, all licences of the contents are propagated forward, enabling the users of the framework to easily determine the allowed uses for any given KG.

Particularly for the creation of databases made available to the public we recommend using the "strict mode" of BioCypher, which does not allow creation of entities without associated source, licence, and version parameters. In this scenario, BioCypher can effectively prevent the re-distribution of data whose original licence does not allow it, and guarantees that data originators are acknowledged.

BioCypher is a free software under MIT licence, openly developed and available at https://biocypher.org. Community contributions in the form of GitHub issues or pull requests are very welcome and encouraged.

## Usage

A key advantage of the modular structure of BioCypher is the ability to reuse existing adapters for primary or secondary knowledge sources. In case no adapter exists for a given resource, it can be created following the pattern of one of our existing adapters and shared with the community for further use. To create a custom KG with BioCypher, two main components are necessary: 1) a YAML file (https://yaml.org) detailing the configuration of graph constituents, including their mode of representation (node or edge) and their preferred identifier; and 2) one or multiple adapter modules responsible for handing off the data to BioCypher. These two components are described below. BioCypher provides a number of utilities for manipulating the input data as well as the ontological foundation of the graph, for instance, filtering properties of input types or arbitrarily extending the ontology. More details and a tutorial can be found in the documentation at https://biocypher.org.

## Schema configuration

Configuration of graph constituents is made available through a graph schema YAML file, whose main purpose is to mediate between the structure of the input data and the resulting BioCypher KG structure. It details, for each constituent class of the graph: its mode of representation (node or edge), the identifier namespace used to identify unique entities (e.g., Ensembl accession or HGNC Symbol for genes), the label to be expected in the input, and - in the case of relationships - the types of source and target nodes. It can also be used to unify the properties attached to nodes and edges in the resulting KG, which is useful when combining sources or dealing with heterogeneous datasets, and to modify and extend the underlying ontology.

## Data retrieval (the adapter)

A Python adapter module is responsible for the actual database creation process. Briefly, the primary data are ingested and passed into BioCypher through the driver instance created in the build pipeline. BioCypher accepts lists and generators; the latter enable streaming of very large datasets that may not fit into the working memory of smaller machines. We provide information about the structure of the input data through the package and its documentation. BioCypher enables automatic extended labelling of each node with the entire hierarchy of that node derived from the ontology tree, which allows more flexibility in querying the resulting KG and simplifies the queries. For instance, a narrow query could yield interactions of proteins, while a query for "polypeptides" (the ontological parent of "protein") yields proteins, peptides, and precursors; a query for "gene or gene product" additionally returns genes and transcripts without the need for concatenating all individual classes of entities or modifying the underlying graph.

## Acknowledgements

This project has received funding from the European Union's Horizon 2020 research and innovation programme under grant agreement No 965193 for DECIDER and No 116030 for TransQST, and the German Federal Ministry of Education and Research (BMBF, Computational Life Sciences grant No 031L0181B and MSCoreSys research initiative research core SMART-CARE 031L0212A).

CTH and BMG were funded under the Defense Advanced Research Projects Agency (DARPA) Young Faculty Award [W911NF-20-1-0255].

JAHW is funded by the BMBF as part of the MIRACUM consortium within the Medical Informatics Initiative Germany (FKZ: 01ZZ2019).

We are thankful to Henning Hermjakob, Benjamin Haibe-Kains, Pablo Rodriguez-Mier, Daniel Dimitrov, and Olga Ivanova for feedback on the initial draft of the manuscript.

## Author contributions

SL conceptualised and developed BioCypher, organised and implemented the case studies, and wrote the manuscript. PA, AFT and EPL contributed to the embedding case study. JB, KD, MH, CK, AM, NP, and JAHW contributed to the federated learning case study. BB and TK contributed to the upscaling case study. PC, FP, and MP contributed to the data integration case study. TD, BSen, and EU contributed to the modularity case study. JD and BSch contributed to the tumour board case study. ID and DO contributed to the Open Targets case study. BMG and CTH contributed to Bioregistry integration. MM and MTS contributed to the subgraph extraction case study. DT contributed to BioCypher development and manuscript writing. JSR conceptualised BioCypher, supervised the project, and acquired funding. All co-authors revised the manuscript.

## Conflict of interest

JSR reports funding from GSK, Pfizer and Sanofi and fees from Travere Therapeutics and Astex Pharmaceuticals.

**Supplementary Materials**

## Case studies

In the following sections, we illustrate the usefulness of various design aspects of BioCypher in practical examples. For most of these case studies, an actual implementation already exists, while some are still drafts or work in progress in early stages. Practical implementations including public code can be accessed for **Modularity**, **Tumour board**, **Network expansion**, **Subgraph extraction**, **Embedding**, and **Open Targets**.

### Modularity

There are several resources used by the biomedical community that can be considered essential to most bioinformatics tasks. A good example is the curation effort on proteins done by the members of the Universal Protein Resource (UniProt) consortium [1]; many secondary resources and tools depend on consistent and comprehensive annotations of the major actors in molecular biology. As such, there are an enormous number of individual tools and resources that make requests to the public interface of the UniProt service, all of which need to be individually maintained. We and several of our close collaborators make use of this resource, for instance in OmniPath [2], CKG [3], Bioteque [4], and the CROssBAR drug discovery and repurposing database [5]. We have created an example on how to share a UniProt adapter between resources and how to use BioCypher to combine pre-existing databases based on ontology.

We have written such an adapter for UniProt data, using software infrastructure provided by the OmniPath backend PyPath (for downloading and locally caching the data). The adapter provides the data as well as convenient access points and an overview of the available property fields using Python Enum classes, offering automatic suggestion and autocomplete functionality. Using these methods, selecting specific content from the entirety of UniProt data and integrating this content with other resources is greatly facilitated (**Figure S1**), since the alternative would be, in many cases, to use a manual script to access the UniProt API and rely on manual harmonisation with other datasets.

Similarly, we have added adapters for protein-protein interactions (PPI) from the popular sources IntAct [6], BioGRID [7], and STRING [8]. By using the UniProt accession of proteins in the KG and BioCypher functionality, the sources are seamlessly integrated into the final KG despite their differences in original data representation. As with UniProt data, access to PPI data is facilitated by provision of Enum classes for the various fields in the original data. The adapters and a script demonstrating their usage are available at https://github.com/HUBioDataLab/CROssBAR-BioCypher-Migration.

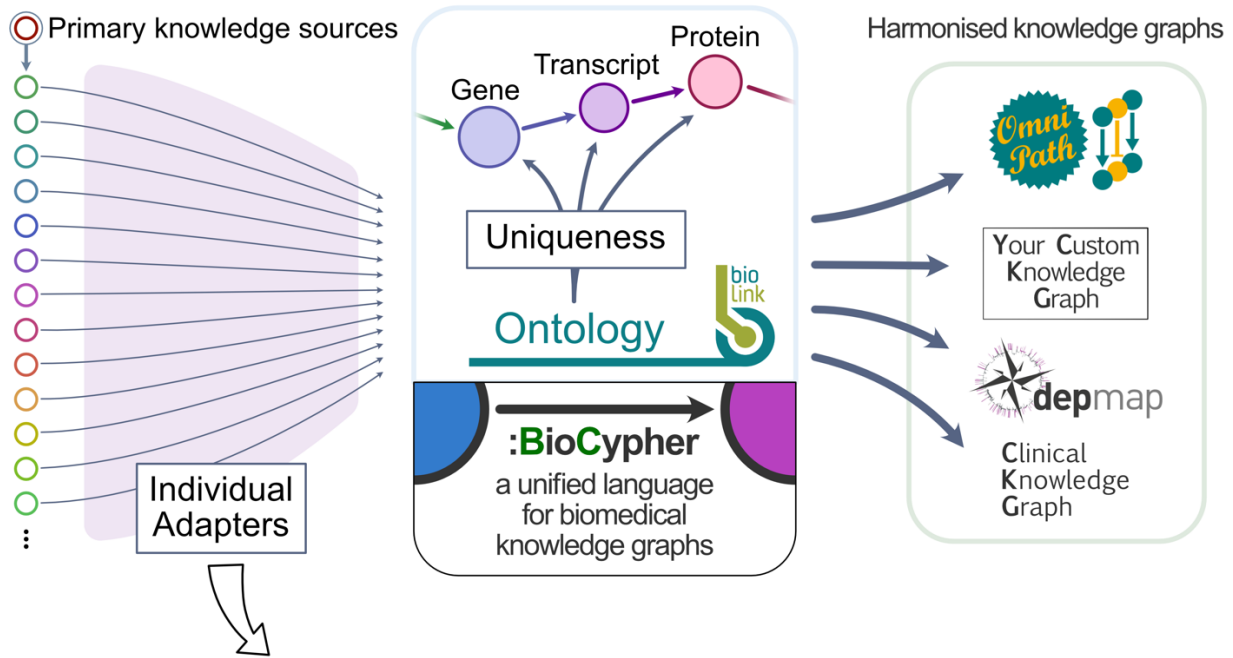

*Figure S1*: *Modularity of knowledge input*. Individual primary source adapters can be used to build secondary knowledge curations such as OmniPath (compare to Figure 1A). This shifts maintenance towards the primary source and thus reduces maintenance effort: instead of maintaining each primary resource at the integrated KG level, only one reusable adapter for each resource is necessary. The primary adapters provide an additional level of flexibility to the user by providing accessible insight into the contents of each primary resource, which can be extensive. For instance, in the adapter for the UniProt knowledge base, the user can select their favourite species, fields of protein information such as the length or mass of the protein, and relationships to import, such as the host organism or the coding gene of each protein.

## Tumour board

Cancer patients nowadays benefit from a large range of molecular markers that can be used to establish precise prognoses and direct treatment [9,10]. In the context of the DECIDER project (www.deciderproject.eu), we are creating a platform to inform the tumour board of actionable molecular phenotypes of high-grade serous ovarian cancer patients. The current manual workflow for discovering actionable genetic variants consists of multiple complex database queries to different established cancer genetics databases [10–12]. The returns from each of the individual queries then need to be curated by human experts (geneticists) in regard to their identity (e.g. identify duplicate hits from different databases), biological relevance, level of evidence, and actionability. The heterogeneous nature of results received from different primary database providers makes this a time-consuming task, and a bottleneck for the discovery and comprehensive evaluation of all possible treatment options.

To facilitate the discovery of actionable variants and reduce the manual labour of human experts, we use BioCypher to transform the individual primary resources into an integrated, task-specific KG. Through mapping of the contents of each primary resource to ontological classes in the build process, we largely remove the need to manually curate and harmonise the individual database results. This mapping is determined once, at the beginning of the integration process, and results in a BioCypher schema configuration that details the types of entities in the graph (e.g., patients, different types of variants, related treatment options, etc.) and how they are mapped and thus integrated into the underlying ontological framework. As a second step, datasets that are not yet available from pre-existing BioCypher adapters are adapted in similar fashion to yield data ready to be ingested by BioCypher. The code for this project can be found at https://github.com/oncodash/oncodashkb.

We make use of the ontology manipulation facilities provided by BioCypher to extend the broad but basic Biolink ontology in certain branches where it is useful to have more granular information about the data that enters the KG. For example, the exact type of genetic variants are of high importance in the molecular tumour board process, but Biolink only provides a generic "sequence variant" class in its schema. Therefore, we extended the ontology tree at this node with the very granular corresponding subtree of the Sequence Ontology (SO, [13]), yielding a hybrid ontology with the generality of Biolink and the accuracy of a specialised ontology of sequence variants (**Figure S2**). Building on the mechanism provided by BioCypher, this hybridisation can be performed by providing only the minimal input of the sequence ontology URL and the nodes that should be the point of merging ("sequence variant" in Biolink and "sequence_variant" in SO). The same process is used with the Disease Ontology [14] and OncoTree [15] (see **Figure S2**).

Once the database has been created through BioCypher, the process of querying for an actionable variant and its associated treatment options for a given patient is greatly simplified. This approach also improves the concordance of knowledge base sources, the ability to incorporate external clinical resources, and the recovery of evidence only represented in a single resource [10].

The major advantage of using BioCypher to integrate several resources is the formal representation of the process provided by the schema configuration, which allows for a simple description and long-term centralised maintenance. Other approaches [10] need ad-hoc scripts, hindering refactoring if the input resources change, and lose metadata about the provenance of the merged information, hindering *a posteriori* analysis.

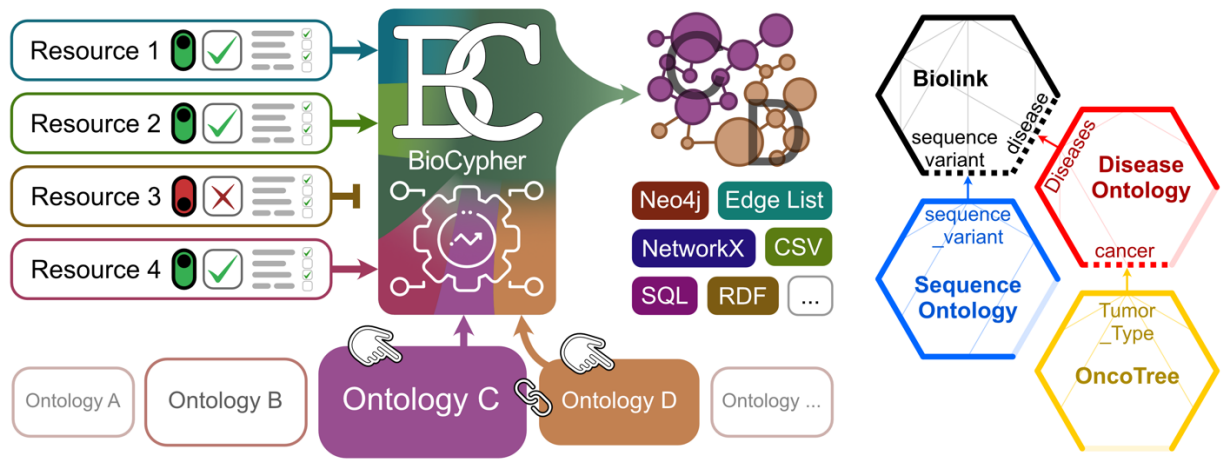

***Figure S2: Modular ontology.*** *BioCypher combines modular inputs from biomedical resources (left) with a flexible scaffold based on ontology (bottom) to build task-specific knowledge graphs (KGs) with variable format (middle). Users can configure the use of individual resources as well as the contents taken from these resources in a community-curated collection of adapters. The data are then harmonised on the basis of user-specified ontologies that are tailored to the specific purpose of the desired KG, using BioCypher's mapping, extension, and hybridisation facilities. Finally, the KG is provided to the user through an output adapter in the desired format. Since Biolink has a broad but general representation of biomedical classes, we extend the "sequence variant" with the corresponding granular information from the specialised Sequence Ontology (right side). Similarly, information about cancer and specific tumour types are added from Disease Ontology and OncoTree.*

## Network expansion

Database schemata of large-scale biomedical knowledge providers are tuned for effective storage. For analysis, the user may benefit from a more dedicated schema type corresponding to the biological question under investigation. We created BioCypher with the objective to simplify the transformation from storage-optimised schemas to analysis-focused schemas. Given one or multiple data sources, the user should be able to quickly build a task-specific knowledge graph using only a simple configuration of the desired graph contents. We demonstrate the simplifying capabilities using an interaction-focussed graph database derived from the Open Targets platform as an example [16].

Barrio-Hernandez et al. used this graph database to inform their method of network expansion [17]. The database runs on Neo4j, containing about 9 million nodes and 43 million edges. It focuses on interactions between biomedical agents such as proteins, DNA/RNA, and small molecules. Returning one particular interaction from the graph requires a Cypher query of ~13 lines which returns ~15 nodes with ~25 edges (variable depending on the amount of information on each interaction). A procedure to collect information about these interactions from the graph is provided with the original manuscript [17], containing Cypher query code of almost 400 lines (http://ftp.ebi.ac.uk/pub/databases/intact/various/ot_graphdb/current/apoc_procedures_ot_data.txt). Still, this extensive query only covers 11 of the 37 source labels, 10 of the 43 target labels, and 24 of the 76 relationship labels that are used in the graph database, offering a large margin for optimisation in creating a task-specific KG.

After BioCypher adaptation, the KG (covering all information used by Barrio-Hernandez et al.) has been reduced to ~700k nodes and 2.6 million edges, a more than ten-fold reduction, without loss of information with regard to this specific task. Compared to the original file of the database dump (zipped, 1.1 GB), the BioCypher output is ~20-fold smaller (zipped, 63 MB), which greatly facilitates sharing and accessibility (e.g. by simplifying online access via Jupyter notebooks). The Cypher query for an interaction has been reduced from 13 query lines, 15 nodes, and 25 edges to 2 query lines, 3 nodes, and 2 edges (**Figure S3**).

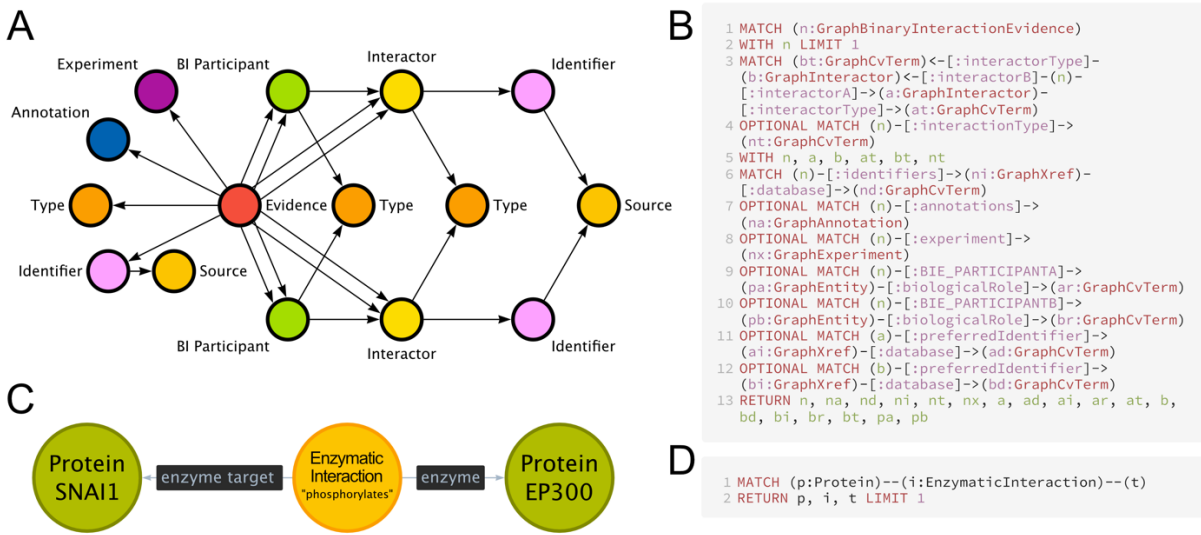

***Figure S3: Semantic abstraction.** A) The original, "storage-oriented" format used by the OTAR KG, displaying one interaction with additional data. B) The Cypher query to receive one interaction from the OTAR graph. C) The migrated, "task-oriented" format produced by the BioCypher adapter, displaying one interaction. The "additional data" from (A) about experiment and evidence type can be added to the interaction node as a property or encoded in additional nodes connected to the interaction node. D) The Cypher query to receive one interaction from the migrated graph.*

Most of this reduction is due to removal of information that is not relevant to the task at hand and semantic abstraction; for instance, the original chain of `("hgnc")-[:database]-("SNAI1")-[:preferredIdentifier]-(:Interactor)-[:interactorB]-(:Interaction)-[:interactorA]-(:Interactor)-[:preferredIdentifier]-("EP300")-[:database]-("hgnc")` to qualify one protein-protein-interaction can be reduced to `("EP300")-[:enzyme]-("phosphorylation")-[:enzyme target]-("SNAI1")`. Arguably, the shorter BioCypher query is also more informative, since it details the type of interaction as well as the roles of the participants. In addition, this representation returns sources of information about the proteins and the interaction as properties on the nodes, and the hierarchical ontology-derived labels provide rich information about the biological context. For instance, the first ancestor labels of the "*phosphorylation*" node are "*enzymatic interaction*", "*direct interaction*", and "*physical association*", grounding this specific interaction in its biological context and enabling flexible queries for broader or more specific terms. This additional information was introduced into the data model by combining the Biolink ontology with the molecular interaction ontology by the Proteomics Standards Initiative [18]. Thus, this "task-oriented" representation is complementary to the "storage-oriented" one, serving a different purpose, and BioCypher provides an easy and reliable way of going from one type of representation to the other.

The BioCypher migration is fast (about 15 minutes on a common laptop) and tested end-to-end, including deduplication of entities and relationships as well as verbose information on violations of the desired structure (e.g., due to inconsistencies in the input data), making the user explicitly aware of any fault points. Through this feedback, several inconsistencies were found in the original Open Targets graph during the migration, some of which originated from misannotation in the SIGNOR primary resource (e.g., "*P0C6X7_PRO_0000037309*" and "*P17861_P17861-2*"). This problem affected only a few proteins, which could have gone unnoticed in a manual curation of the data; a type of problem that likely is common in current collections of biomedical knowledge.

Knowledge representations can and should be tuned according to the specific needs of the downstream task to be performed; BioCypher is designed to accommodate arbitrarily simple or complex representations while retaining information important to biomedical research tasks. A compressed structure is important, for instance, in graph machine learning and embedding tasks, where each additional relationship exponentially increases computational effort for message passing and embedding techniques [4,19]. Most importantly, evidence (which experiment and publication the knowledge is derived from) and provenance (who provided which aspects of the primary data) should always be propagated. The former is essential to enable accurate confidence measures; e.g., not double-counting the same information because it was derived from two secondary sources which refer to the same original publication. The latter is important for attribution of work that the primary maintainers of large collections of biomedical knowledge provide to the community. The code of this migration can be found at https://github.com/saezlab/OTAR-BioCypher.

**Subgraph extraction**

For many practical tasks in the workflow of a research scientist, the full KG is not required. For this reason, building complex and extensive KGs such as the CKG [3] or the Bioteque [4] would not be sensible in all use cases.

For instance, in the context of a proteomics analysis, the user would only like to contextualise their list of differentially abundant proteins using literature connections in the CKG, rendering much of the information on genetics and clinical parameters unnecessary. In addition, the KG may contain sensitive data on previous projects or patient samples, which cannot be shared (e.g. in the case of publishing the analysis), causing reproducibility issues. Likewise, some datasets cannot be shared due to their licences. With BioCypher, a subset of the entire knowledge collection can be quickly and easily created, taking care to not include sensitive, irrelevant, or unlicensed data. The analyst merely needs to select the relevant species (e.g. proteins, diseases, and articles) and their relationships in the BioCypher configuration.

BioCypher then queries the original KG and extracts the required knowledge, conserving all provenance information, and yielding a much reduced data set ready for sharing.

The original CKG is shared as a Neo4j database dump with a compressed size of 5-7 GB (depending on the version), including 15M nodes and 188M edges. After BioCypher migration of the full CKG, the same KG can be created from BioCypher output files that have a compressed size of 1.3 GB. Of note, the creation from BioCypher files using the admin import command is Neo4j version-independent, which is not the case for dump files and can be a reproducibility issue for earlier versions; for instance, the graph of Barrio-Hernandez et al. in the **"Network expansion"** case study is a Neo4j v3 dump, which is no longer supported by the current Neo4j Desktop application. Finally, after the subsetting procedure, the reduced KG (including 5M nodes and 50M edges) in BioCypher format has a compressed size of 333 MB.

Since a complete CKG adapter already existed (found at https://github.com/saezlab/CKG-BioCypher/), the subsetting required minimal effort; i.e., the only required step was to remove unwanted contents from the complete schema configuration. The code for this task can be found at https://github.com/saezlab/CKG-BioCypher/tree/subset.

### Embedding

As a second subsetting example, we demonstrate the usefulness of subsetting KGs for task-specific graph embeddings. KG embeddings can be an efficient lower-dimensional substitute for the original data in many machine learning tasks [4] and, as methods such as GEARS [20] show, these embeddings can be useful for very complex, hard tasks. However, including all prior data in every embedding is not necessary for good results, while using the proper domain of knowledge can vastly increase the performance of downstream tasks [4]. This issue extends both to the type of knowledge represented (not every kind of relationship is relevant to any given task) as well as the source of the knowledge (different focus points in knowledge resources lead to differential performance across different tasks). Thus, it is highly desirable to have a means to identify the proper knowledge domain relevant to a specific task to increase the efficiency of subsequent analyses.

To achieve this aim, BioCypher can facilitate task-specific builds of well-defined sets of knowledge from a combination of primary sources for each application scenario. And, since the BioCypher framework automates much of the build process going from only a simple configuration file, the knowledge representations can be iterated over quickly to identify the most pertinent ones. As above, the only requirement from the user (given existing BioCypher adapters for all requested primary sources) is a selection of biological entities and their relationships in the schema configuration.

We have performed this method of subsetting embedding in the Bioteque environment [4] with a subset of the Clinical Knowledge Graph [3]. Concretely, we emulated a scenario where a user seeks to computationally describe the patient samples available in the CKG as a means to explore context-specific similarity between patients. In brief, we first selected a few sequences of relationships (i.e. the metapath) to connect subjects (patients) to the proteins expressed by their individual samples, (i.e. subject → biological sample → analytical sample → protein). Given the rich variety of associations available for protein entities, we can further link these subjects to other entities and relations available in the knowledge graph, enabling the exploration of specific contexts. For instance, we extended the metapath to connect the subjects' protein readouts to biological pathways. Importantly, due to the gigantic size of the CKG, it was fundamental to use a CKG BioCypher adapter to extract the pertinent subgraphs containing only the required knowledge (e.g. patient-protein data and pathways). Indeed, selecting the desired KG entities from the complete adapter required negligible time (demonstrated at https://github.com/saezlab/CKG-BioCypher). Finally, the protein- and pathway-based patient descriptors were obtained by running the Bioteque embedding pipeline (https://gitlabsbnb.irbbarcelona.org/bioteque/). The two resulting patient embedding spaces and their corresponding cluster similarity are provided in **Figure S4**.

Note that, thanks to the modular nature of the Bioteque pipeline, it is possible to generate embeddings from any network (even beyond the ones used in the Bioteque KG) by just extracting the connections forming the metapath. In this regard, BioCypher offers a handy means to query the pertinent input files for the Bioteque pipeline, paving the way for an efficient exploration, identification, and extraction of task-specific KG contexts (e.g. generation of KG embeddings for patient similarity exploration). Indeed, a similar exercise can be performed on the Open Targets dataset (see next section), with minimal preparatory effort. This would allow, for instance, to further connect protein readouts to disease associations or to complement patient descriptors with embeddings of diseases, drugs and drug targets for downstream predictive pipelines.

**Open Targets**

The Open Targets platform is an open resource for drug discovery provided by the European Bioinformatics and Sanger Institutes [16]. Their core dataset on drug target-disease relationships is provided for download in columnar format; it is internally harmonised but only partially mapped to several disjoint ontologies (mainly disease-related). The dataset can be downloaded in Parquet format, a data structure designed to work on distributed systems in a highly parallel manner, making efficient BioCypher adaptation very simple.

To enable an open, community-maintained KG version of the columnar Open Targets dataset, we created a BioCypher adapter (https://github.com/saezlab/OTAR-BioCypher). Due to the efficient data processing using Parquet and PySpark, the adapter can be run on small machines such as current laptops as well as in distributed high-performance computing environments. It provides a flexible basis for individually customised KGs from Open Targets data and allows frequent rebuilding of the KGs when the dataset is updated. The simple layout of a BioCypher adapter allows rapid implementation (less than 500 lines of code) and response to breaking changes in the source material (such as structural or name changes). Additionally, since the adapter can be reused, changes need to be implemented only once for the benefit of all downstream users.

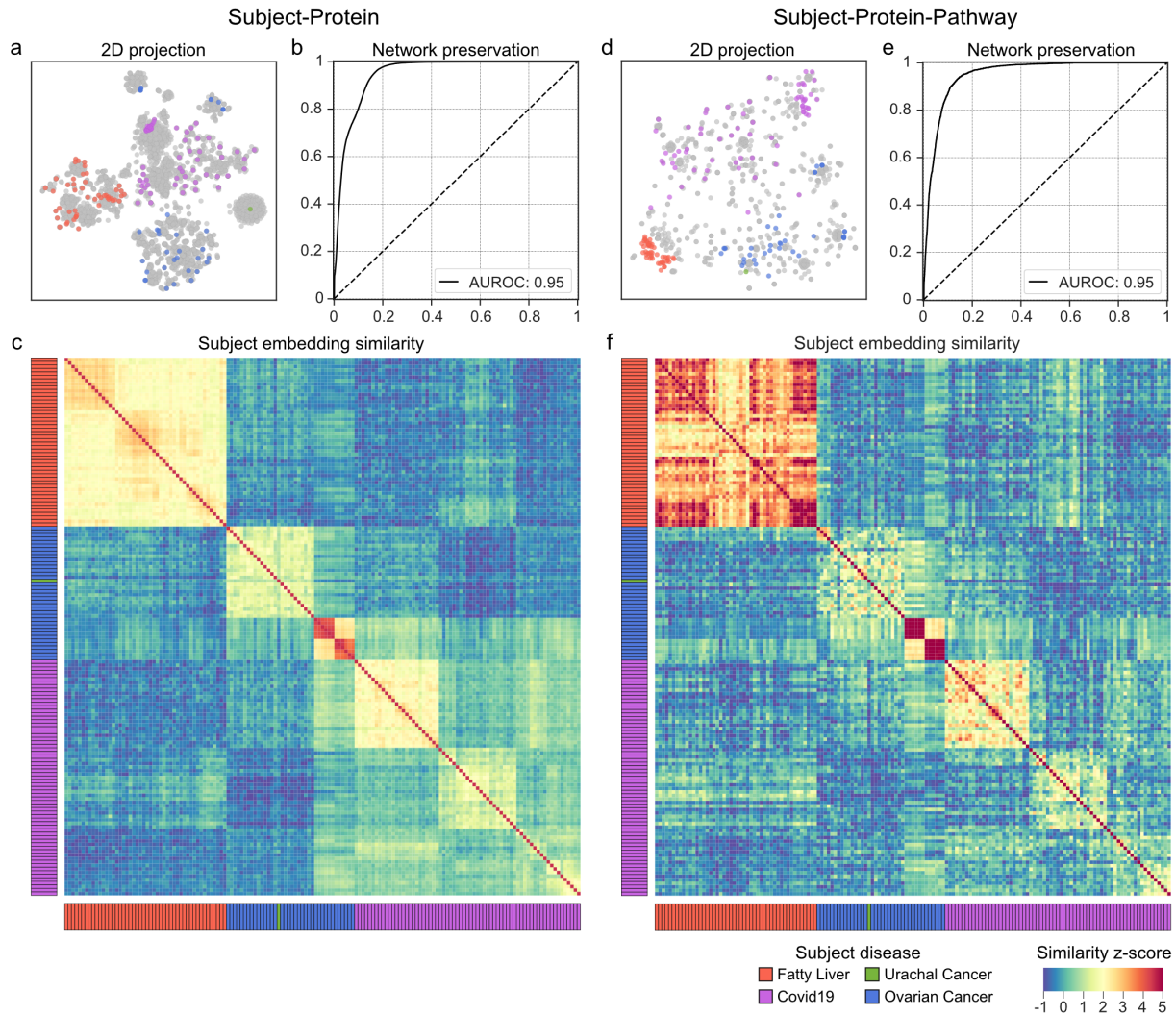

***Figure S4: Bioteque-based patient embeddings.*** *Two embedding spaces were obtained to describe patients ('Subjects' nodes in the CKG) based on protein (left) and protein-pathway (right) similarities. a, d: 2D projection (t-SNE) of the subjects according to the protein similarity (a) and pathway similarity (d). Subject nodes are coloured by disease type (see legend at the bottom) while proteins (a) and pathways (d) are coloured in grey. b, e: ROC curve validation when reconstructing the original networks with the corresponding embeddings. c, f: Heatmap showing the cosine similarity of the subject embeddings. To make similarities comparable between heatmaps, cosine similarities were transformed into z-scores by subtracting the median and dividing by the interquartile range of their corresponding background distribution. Thus, the higher/redder the z-score the higher the similarity. An agglomerative hierarchical clustering based on the protein-driven similarities (c) was used to sort the rows and columns in both heatmaps (c,f). Rows (columns) are coloured based on the subject's disease (see legend at the bottom). Notice how, while both spaces are unsurprisingly similar (i.e. both are based on protein readouts), the pathway-driven similarities reveal sub-clusters within disease types that were not evident based on purely protein-driven similarities.*

As shown in the case study "Modularity", user access of the data is facilitated by Enum classes detailing the dataset contents, allowing automatic suggestions and autocomplete, including all individual source datasets. Licences of all original data are propagated, and the use of BioCypher "strict mode" guarantees the inclusion of *licence*, *source*, and *version* fields on every single entity of the KG, greatly simplifying downstream decisions related to licensing.

Mapping the Open Targets dataset to a central ontology also facilitates integration with further datasets such as UniProt and the Cancer Dependency Map. Since Open Targets is a gene-centric platform, data from UniProt can yield complementary insights on the protein layer, for instance by coupling to other datasets of signalling cascades. We included information on human proteins by simply adding the protein node type and the gene-to-protein edge from the UniProt adapter described in section **Modularity**. Harmonising the data was then a simple matter of loading the additional adapter, making sure that the identifier namespace used for genes (ENSEMBL gene) was the same in both adapters (via Enum-based configuration), and writing the information to disk via BioCypher. It only required the addition of 8 lines of code in the build script. Adding gene essentiality and cell line information from the Dependency Map project adapter was performed similarly by adding the adapter and loading nodes and edges in the correct format.

### Federated learning

Federated learning is a machine learning approach that enables multiple parties to collaboratively train a shared model while keeping their data decentralised and private [21,22]. This is achieved by allowing each party to train a local version of the model on their own data, and then sharing the updated model parameters with a central server that aggregates these updates. However, most machine learning algorithms depend on a unified structure of the input; when it comes to algorithms that combine prior knowledge with patient data, a large amount of harmonisation needs to occur before the algorithms can be applied.

BioCypher facilitates federated machine learning by providing an unambiguous blueprint for the process of mapping input data to ontology. Once a schema for a specific machine learning task has been decided on by the organisers, the BioCypher schema configuration can be distributed, ensuring the same database layout in all training instances. The usefulness becomes apparent in two pilot projects outlined below.

Firstly, the Care-for-Rare project of the Munich Children's Hospital has to synchronise a broad spectrum of biomedical data: demographics, medical history, medical diagnosis, laboratory results from routine diagnostics, imaging and omics data with analyses of proteome, metabolome and transcriptome in different tissues as well as genetic information. To allow reaching a sample size that is suitable for modern methods of diagnosis and treatment options in rare diseases [23], world-wide collaboration between children's hospitals is a necessity. The unstructured nature of most clinical data necessitates a harmonisation step with subtle challenges with respect to ontology. For instance, general classifications such as ICD10-GM subsume rare childrens' diseases under umbrella terms for whole disease groups, requiring alternative coding catalogues such as Orphanet OrphaCodes [24] and the German Alpha-ID [25].

Larger ontologies such as HPO [26] and SNOMED-CT [27] are complex and expanded constantly. In addition to the technical challenges, the legal requirements of patient confidentiality and data protection necessitate extreme care in the processing of all data, hindering information sharing between collaborators. All of the above poses great challenges in data integration in the clinical setting.

Secondly, the MeDaX project (bio**Me**dical **Da**ta e**X**ploration at University Medicine Greifswald) develops innovative and efficient methods for storage, enrichment, comparison, and retrieval of biomedical data based on KG technology. Embedded in the Medical Informatics Initiative (MII) Germany, MeDaX builds on the federated storage structure for biomedical health care and research data established in all Data Integration Centres (DICs) at German university hospitals. We envision extending the existing MIRACOLIX toolbox [28] with the MeDaX pipeline to set up local KGs, combining complex heterogeneous data from multiple resources: in addition to biomedical data available only at the DICs due to patient privacy, we include the MII core data set [29], local population studies [30,31], biomedical ontologies [32], and public information portals [33]. BioCypher's ontology mapping process facilitates future integration of additional data sources (see also the **case study "Data integration"**).

We enable federated learning pipelines by supplying build instructions for each local database in the form of the schema configuration that can be publicly and centrally maintained, since it contains no sensitive data (**Figure S5**). At each training location, a task-specific KG is created from public data (e.g., with the Clinical Knowledge Graph as baseline), using the subsetting facilities described in the **case study "Subgraph extraction"**. Afterwards, the sensitive patient data (e.g., germ-line genetic variants) are integrated into this KG at each location, using the BioCypher schema configuration to specify the type of data involved (e.g., clinical measurements, genetic profiling). This ensures that, regardless of how the sensitive data are represented at each location, the machine learning algorithm works with the exact same structure of KG, preventing accidental or malicious data leakage in the federated learning step.

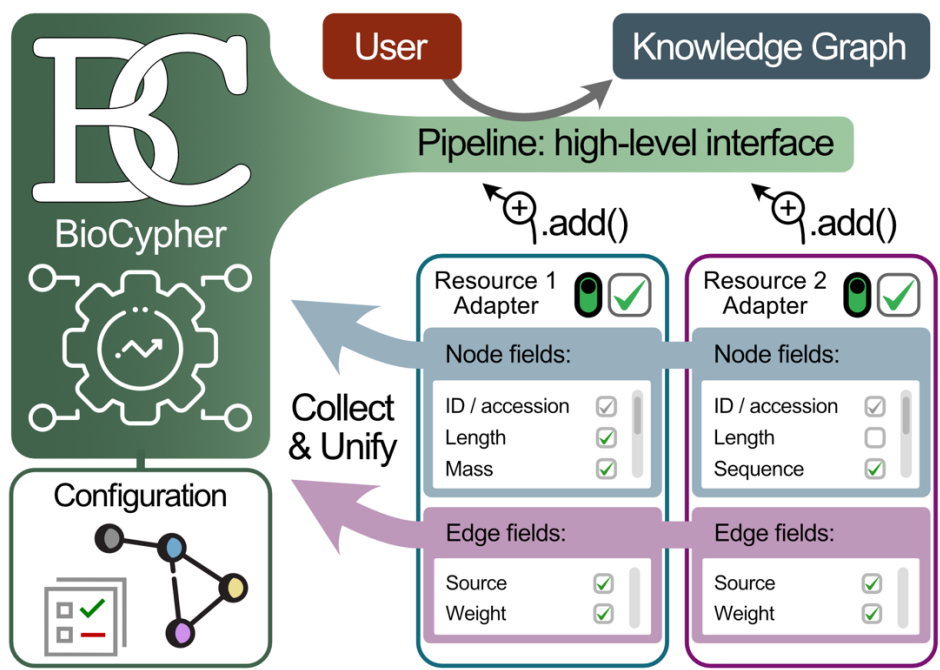

***Figure S5: User interface.*** *BioCypher provides high-level users with an abstracted pipeline interface that is used to aggregate data from primary adapters while collecting and unifying the individual data inputs. Configuration needs to take place only globally when combining adapters that provide overlapping identifier systems, which can be assessed through the pipeline interface. This is useful to synchronise proprietary or sensitive data between single locations in a federated learning pipeline, since the adapters that contain non-public data only need to provide non-sensitive, summary level information about the data they supply.*

## Data integration

Biomedical data collections are growing to enormous sizes, which makes the handling of data alone a non-trivial task. Additionally, these large corpuses then need to be put to good use in downstream analyses, including collaborations between groups or even institutions. The growth of arbitrarily organised large-scale collections of knowledge poses major challenges to the maintainers of these databases:

1. Maintaining data ingestion pipelines for dozens of upstream data sources is not feasible in a research context and detracts development time from other tasks.
2. Using a custom (non-standardised) data model, the effort to integrate new upstream data sources grows with the total number of pre-existing data sources. Each new data source has to be cross-referenced with all existing data sources and inconsistencies arise because the same piece of information may be represented with different levels of abstraction.
3. The custom data model also complicates collaboration with external researchers. Integrating data from different contexts requires the collaborators to adapt to the internal data model.

BioCypher can handle all three challenges. Firstly, the open architecture and community effort around BioCypher allows maintaining core data ingestion pipelines while reusing data adapters from experts in other fields. Secondly, the well-described data model of Biolink drastically reduces the effort required to integrate new data sources because they need only to be adapted to the core data model, not to all existing data. Thirdly, the combination of an open architecture and ontology-based data integration facilitates collaborations with external researchers. We maintain two pilot projects for continuous large-scale data integration in a research context, detailed below.

1) The German Centre for Diabetes Research (DZD, www.dzd-ev.de) has developed a knowledge graph to support data integration for translational research. The internal KG instance provided the foundation of the open-source CovidGraph project [34] which is now maintained by the HealthECCO community (www.healthecco.org). At the core of the DZD KG is a data ingestion pipeline for PubMed that transforms publication data into a detailed graph representation, including authors, affiliations, references, and MeSH term annotations. The PubMed graph contains 350 million nodes and 850 million relationships, as well as data on biological entities (genes, transcripts, proteins), their functional annotations, and biochemical interactions. This KG is used to link internal research data to public knowledge and to generate new research hypotheses.

Re-building the data ingestion and maintenance based on BioCypher reduces the time required to bring new data products to researchers at the DZD because the unified data model and ontology-backed data harmonisation allow the reuse of data analysis modules and user interface components. Removing obstacles for collaboration on the knowledge graph supports interdisciplinary research on diabetes complications and comorbidities.

2) At the National Centre for Tumour Diseases (NCT) and the German Cancer Research Centre (DKFZ), we aim to integrate a biomedical knowledge graph with patient data from clinical studies, including multi-omics data, to aid in the stratification of novel biomarkers and the implementation of precision medicine. To achieve this, we are using the BioCypher framework to create a biomedical KG from curated primary data sources. The KG will be expanded over time through experimental results as well as clinical annotation and will provide an interface for different roles in the cancer research process. The maintenance and integration of the biomedical knowledge graph with patient data offers new opportunities for analysis that may enhance the accuracy and effectiveness of precision medicine approaches.

**Upscaling**

As biomedical data become larger, integrated analysis pipelines become more expansive and, thus, expensive. For numerous projects in systems biomedicine to succeed, a flexible way of maintaining and analysing large sets of knowledge is necessary. This is done most effectively by separating data storage and analysis (such that each component can be individually scaled), while using distributed computing infrastructure to perform both tasks in close vicinity, such as computing clusters. We have recently published an open source software, called Sherlock, to perform this type of data management for biomedical projects [35]. However, this pipeline in some ways depends on manual maintenance, for instance in its data transformation from primary resource to internal format (https://github.com/earlham-sherlock/earlham-sherlock.github.io/tree/master/loaders).

Using BioCypher, we facilitate the maintenance of Sherlock's input sources by reusing existing adapters and converting the manual scripts to additional adapters for unrepresented resources. Combined with the unambiguous BioCypher schema configuration, this will make Sherlock's input side automatable and greatly decrease maintenance effort, unlocking its full potential in managing complex bioinformatics projects and their resources. Given a configuration that can be developed locally, a project database can be upscaled to arbitrary numbers of nodes on an in-house or commercial cluster just as the project requires, saving computing time and thereby money. By virtue of the Sherlock-BioCypher integration, these projects retain the benefits from both frameworks; BioCypher provides reusability, transparency, and ontological grounding, while Sherlock makes data storage and analysis vastly more efficient and economical.

**Supplementary table 1. Recent biomedical knowledge graph solutions (non-comprehensive).**

| Database | Reference |
|---|---|
| Biological Insight Knowledge Graph | [36] |
| Bioteque | [4] |
| Clinical Knowledge Graph | [3] |
| CROssBAR | [5] |
| Dependency Map | [37] |
| GenomicKB | [38] |
| HealthECCO Covidgraph | [34] |
| INDRA CogEx | https://github.com/bgyori/indra_cogex |
| KG-COVID-19 | [39] |
| OmniPath | [2] |
| Open Targets | [16] |
| PheKnowLator | [40] |
| PORI (Platform for Oncogenic Reporting and Interpretation) | [10] |
| PrimeKG | [41] |
| RTX-KG2 | [42] |
| TypeDB | https://github.com/typedb-osi/typedb-bio |